\documentclass[12pt,a4j]{article}
\usepackage[dvips]{graphicx}
\usepackage{amsmath}
\usepackage{enumerate}
\begin{document}
\renewcommand{\thefootnote}{\fnsymbol{footnote}}
\baselineskip=7mm
\centerline{\bf The Peakon Limit of the $N$-Soliton Solution of the Camassa-Holm Equation } \par
\bigskip
\centerline{Yoshimasa Matsuno\footnote{{\it E-mail address}: matsuno@yamaguchi-u.ac.jp}}\par

\centerline{\it Division of Applied Mathematical Science,} \par
\centerline{\it Graduate School of Science and Engineering,}\par
\centerline{\it Yamaguchi University, Ube 755-8611} \par
\centerline{(Received    \hspace{5cm})}\par
\bigskip
\bigskip
\par
  We show that the analytic $N$-soliton solution of the Camassa-Holm (CH)
shallow-water model equation converges to the nonanalytic $N$-peakon solution
of the dispersionless  CH equation when the dispersion parameter
tends to zero. To demonstrate this, we develop a novel limiting procedure and
apply it to the parametric representation for the $N$-soliton solution of the
CH equation. In the process, we use Jacobi's formula for determinants as well as
various identities among the Hankel determinants to facilitate
the asymptotic analysis. We also provide a new representation of the $N$-peakon
solution in terms of the Hankel determinants. \par
 \par

 \par
\bigskip
\bigskip
\newpage
\leftline{\bf  1. Introduction} \par
 The Camassa-Holm (CH) equation is a model equation 
describing the unidirectional propagation
of nonlinear shallow-water waves over a flat bottom.$^{1-4)}$  It may be written  in 
a dimensionless  form as
  $$u_t+2\kappa^2 u_x-u_{txx}+3uu_x=2u_xu_{xx}+uu_{xxx}, \eqno(1.1)$$
where $u=u(x, t)$ is the fluid velocity, $\kappa$ is a  positive parameter 
related to the phase velocity of the linear dispersive wave and the subscripts $t$ and $x$
appended to $u$ denote partial differentiation.  Like the Korteweg-de Vries equation
  in shallow-water theory, the CH equation  is a completely integrable nonlinear partial differential  
equation with a rich mathematical structure. Because of this fact, a large number of
works have been devoted to the study of the equation from both physical and 
mathematical points of view.  We do not provide a comprehensive
review on various consequences established for the equation. Instead, we refer to an
excellent survey given in ref. 5). Our main concern here  is solutions to the CH
equation for both $\kappa\not=0$ and $\kappa=0$. Although equation (1.1) can be
transformed to the corresponding equation with $\kappa=0$ by a Galilean
transformation $x^\prime=x+\kappa^2 t, t^\prime=t, u^\prime=u+\kappa^2$, solutions
under the same vanishing boundary condition at infinity have quite different characteristics.
In fact, 
in the case of $\kappa\not=0$, 
it typically exhibits analytic multisoliton solutions (the so-called 
$N$-soliton solution with $N$ being an arbitrary positive integer) like the usual soliton equations.$^{6-15)}$
We also remark that it has singular cusp soliton solutions as well as solutions
consisting of an arbitrary number of solitons and cusp solitons.$^{16-19)}$
  When $\kappa=0$, on the other hand,
eq. (1.1) becomes a dispersionless version of the original CH equation.
It admits a new kind of solitary waves  which have a discontinuous slope at their crest.$^{1,2)}$ 
For this unique feature,  they are now termed $\it peakons$.
Hence, the peakon has a nonanalytic nature unlike the smooth soliton.
One of its remarkable properties is that the motion of peakons is 
described   by a finite dimensional completely integrable dynamical system. This fact was used 
in refs. 20 and 21
 to construct the general $N$-peakon solution which represents the interaction 
of $N$ peakons.
The careful inspection of the interaction of peakons reveals that it occurs elastically in pairs
and the total effect of the collision is the phase shift whose explicit formula
has been given in a closed form.$^{2, 9, 21)}$ See also a recent paper concerning
the detailed investigation of the dynamics of two peakons.$^{22)}$ 
An important issue  about the $N$-peakon solution is  
how one can reduce it from the analytic $N$-soliton
solution by taking the singular limit $\kappa\rightarrow 0$. 
Although this problem has received considerable attention, it has been
resolved only partially.
Indeed, in the case of $N=1$, the convergence of the single soliton to the single peakon
has been demonstrated using the explicit form of the $1$-soliton solution.$^{2,5)}$ 
Also, an analysis without recourse to the explicit form of the soliton solution 
has been carried out in a  general context by employing an abstract theory
of dynamical systems.$^{23,24))}$
However, the treatment of the general $N$-soliton case has remained open.
Quite  recently, the case $N=2$ was completed  by two different ways.
One will be given in ref. 25 while the other is included here as a special version of
the general $N$-soliton case. \par
The purpose of this paper is to demonstrate  the convergence of the $N$-soliton
solution to the $N$-peakon solution for the general $N$ by using the explicit $N$-soliton
solution of the CH equation.
In \S 2, we present a parametric representation for the  $N$-soliton solution of the CH equation which offers a
relevant form to the subsequent asymptotic analysis. 
In \S 3, the convergence
of the $1$-soliton solution to the $1$-peakon solution 
is exemplified to explain the  new idea used in the limiting procedure. 
In \S 4, we perform the corresponding limit for the $N$-soliton solution 
and show that it reproduces the $N$-peakon solution given
by Beals et al.$^{21)}$ 
We also obtain a new representation of the $N$-peakon solution in terms of the Hankel determinants.
 Section 5 is devoted to the concluding remarks. 
In Appendix A, we give a proof of the formula which enables us to rewrite the
tau-functions for the $N$-soliton solution of the CH equation in terms of the
Hankel determinants. In Appendix B, we establish various identities among
the Hankel determinants which are used effectively to simplify the
limiting waveform of the $N$-soliton solution.
\par
\bigskip
\leftline{\bf 2. The $N$-Soliton Solution of the CH Equation} \par
 The $N$-soliton solution of the CH equation (1.1) may be represented in a parametric form$^{11,15)}$
$$u(y,t)=\left({\rm ln}\ {f_2\over f_1}\right)_t, \eqno(2.1a)$$
$$x(y,t)={y\over\kappa}+{\rm ln}\ {f_2\over f_1}+d, \eqno(2.1b)$$
with
$$f_1=\sum_{\mu=0,1}{\rm exp}\left[\sum_{i=1}^N\mu_i(\xi_i+\phi_i)
+\sum_{1\le i<j\le N}\mu_i\mu_j\gamma_{ij}\right], \eqno(2.2a)$$
$$f_2=\sum_{\mu=0,1}{\rm exp}\left[\sum_{i=1}^N\mu_i(\xi_i-\phi_i)
+\sum_{1\le i<j\le N}\mu_i\mu_j\gamma_{ij}\right], \eqno(2.2b)$$
where
$$\xi_i=k_i\left(y-\kappa c_it-y_{i0}\right),
\qquad c_i={2\kappa^2\over 1-\kappa^2k_i^2},
\qquad (i=1, 2, ..., N),
\eqno(2.3a)$$
$$e^{-\phi_i}={1-\kappa k_i\over  1+\kappa k_i},\qquad (0<\kappa k_i<1),
\qquad (i=1, 2, ..., N), \eqno(2.3b)$$
$$e^{\gamma_{ij}}={(k_i-k_j)^2\over
(k_i+k_j)^2}, \qquad (i, j=1, 2, ..., N; i\not=j).
\eqno(2.3c)$$
Here, $k_i$ and $y_{i0}$ are soliton parameters 
characterizing the amplitude and phase of the $i$th soliton, respectively,
$c_i$ is the velocity of the $i$th soliton in the $(x,t)$ coordinate system
and $d$ is an integration constant. 
We assume that $c_i>0$ and $c_i\not=c_j$ for $i\not=j\ (i, j=1, 2, ..., N).$
In (2.2), the notation $\sum_{\mu=0,1}$ implies the summation over all
possible combination of $\mu_1=0, 1, \mu_2=0, 1, ..., \mu_N=0, 1$. 
The following coordinate transformation $(x, t) \rightarrow (y, t^\prime)$ 
has been introduced to parametrize 
the $N$-soliton solution
$$dy=rdx-urdt, \qquad dt^\prime=dt, \eqno(2.4a)$$
where the variable $r$ is defined by the relation
$$r^2=u-u_{xx}+\kappa^2. \eqno(2.4b)$$
Note in (2.1) that the time variable $t^\prime$ is identified with the original
time variable $t$  by virtue of (2.4a). \par
The 1-soliton solution is given by 
$$u(y,t)={2\kappa^2c_1k_1^2\over 1+\kappa^2k_1^2+(1-\kappa^2k_1^2)\cosh \xi_1}, \eqno(2.5a)$$
$$x(y,t)={y\over \kappa}+{\rm ln}\left[{(1-\kappa k_1)e^{\xi_1}+1+\kappa k_1\over
((1+\kappa k_1)e^{\xi_1}+1-\kappa k_1}\right]+d. \eqno(2.5b)$$
It represents a solitary
wave  travelling to the right  with the amplitude $c_1-2\kappa^2$ and  
velocity $c_1$. 
The property of the $1$-soliton solutions has been explored in detail as well as its
peakon limit. See, for example ref. 5.  
For completeness, we write the explicit $1$-peakon solution of equation (1.1) with $\kappa=0$.
It reads
$$u(x,t)=c_1e^{-|x-c_1t-x_{10}|}, \eqno(2.6)$$
where $x_{10}$ represents the initial position of the peakon. In view of the invariance
property of the dispersionless version of the CH equation under the
transformation $x^\prime= x, t^\prime = -t, u^\prime =-u$, it also admits a peakon solution (2.6)
with a negative amplitude. It represents a peakon with depression propagating to the left.
\par
\bigskip
\leftline{\bf 3. The Peakon Limit of the $1$-Soliton Solution}\par
 We first consider the limiting procedure for the $1$-soliton solution. This will be helpful
 to understand the basic idea in developing the
procedure for the general $N$-soliton solution. 
We start with the tau-functions for the $1$-soliton solution which are the
most important constituents in our analysis.
 These are given by (2.2) with $N=1$. Before proceeding to the limit,
 we find it convenient to shift the phase constant $y_{10}$ as
$y_{10}\rightarrow y_{10}+\phi_1/k_1$ or in terms of the
phase variable $\xi_1\rightarrow \xi_1-\phi_1$  so that
$$f_1=1+e^{\xi_1}, \eqno(3.1a)$$
$$f_2=1+\nu_1^2e^{\xi_1}, \eqno(3.1b)$$ 
where we have put $\nu_1= e^{-\phi_1}$.
The  limit $\kappa\rightarrow 0$ is taken in such a way that the amplitudes of the soliton 
and peakon given respectively by $c_1-2\kappa^2$ and $c_1$ coincide and remain finite.$^{5)}$
This can be attained by taking the limit
 $\kappa\rightarrow 0$ with  the soliton velocity $c_1$ in the $(x,t)$
coordinate system being fixed. 
We see from (2.3a) that the appropriate limit of the wavenumber $k_1$ is carried out by taking
$\kappa k_1\rightarrow 1$. The limiting procedure described here may be called the {\it peakon limit}.
It now follows from (2.3) that 
various wave parameters have the following leading-order asymptotics in the peakon limit
$$\kappa k_1 \sim 1-{\kappa^2\over c_1},\qquad \nu_1 \sim {\kappa^2\over 2c_1}={\lambda_1\over 4}\kappa^2,
 \eqno(3.2a)$$
$$e^{\xi_1}=e^{\kappa k_1\left({y\over\kappa}-c_1t-x_{10}\right)} \sim {f_1\over f_2}e^{x-c_1t-x_{10}}, \eqno(3.2b)$$
where $\lambda_1=2/ c_1$ and $x_{10}=y_{10}/\kappa$. In passing to the last line of (3.2b),
 we have used (2.1b) to eliminate the $y$ variable
and the constant $d$ has been absorbed in the phase constant $x_{10}$. Substituting (3.2b) into (3.1), the
leading terms of $f_1$ and $f_2$ are  found to be as
$$f_1 \sim 1+{f_1\over f_2}z_1, \eqno(3.3a)$$
$$f_2 \sim 1+\epsilon^2\lambda_1^2{f_1\over f_2}z_1, \eqno(3.3b)$$
where we have put 
$$z_1=e^{x-c_1t-x_{10}}, \qquad \epsilon={\kappa^2\over 4}, \eqno(3.3c)$$
for simplicity. If we introduce the new quantity $f$ by $f=f_2/f_1$, we can deduce from (3.3) that
$$f \sim {f+\epsilon^2\lambda_1^2z_1\over f+z_1}. \eqno(3.4)$$
The expression (3.4) yields the following quadratic equation for $f$:
$$f^2+(z_1-1)f-\epsilon^2\lambda_1^2z_1+O(\epsilon^3)=0. \eqno(3.5)$$
If we solve this equation, we can express $f$ as a function of $x$ and $t$. 
At this instant, it is crucial to observe that $f>0$ since 
both $f_1$ and $f_2$ are positive quantities. A positive solution is then
substituted into (2.1a) to obtain $u$. In performing the differentiation, however, we must
replace the $t$ derivative $\partial/\partial t$ by $\partial/\partial t+u\partial/\partial x$
in accordance with the coordinate transformation (2.4a). With this notice in mind, we can rewrite (2.1a) as
$$u=\left({\partial\over\partial t}+u{\partial\over\partial x}\right){\rm ln}\ f. \eqno(3.6)$$
Solving (3.6) with respect to $u$ to express it in terms of a single variable $f$ and its derivatives, we have 
$$u={f_t\over f-f_x}. \eqno(3.7)$$
The derivatives $f_t$ and $f_x$ in (3.7) are obtained simply if one differentiates (3.5) by $t$ and $x$,
respectively. 
To be more specific, as $\epsilon\rightarrow 0$
$$f_t \sim {c_1z_1f-\epsilon^2c_1\lambda_1^2z_1\over 2f+z_1-1},\eqno(3.8a)$$
$$f_x \sim {-z_1f+\epsilon^2\lambda_1^2z_1\over 2f+z_1-1}, \eqno(3.8b)$$
where we have used the relations $ z_{1,t}=-c_1z_1$ 
and $ z_{1,x}=z_1$ which are derived from (3.3c).
After a manipulation using (3.5) and (3.8), we arrive at the expression of $u$ in terms of $f$:
$$u \sim {-c_1f+c_1\over f+z_1}. \eqno(3.9)$$
The final step of the limiting process is to solve (3.5) under the condition $f>0$ and then 
take the limit $\kappa\rightarrow 0$ after substituting
a positive solution  into (3.9). Although the analytical expression is obtained for $f$ by quadrature, we need
only the series solution. This fact is crucial in developing the peakon limit
of the general $N$-soliton solution where  equation corresponding to (3.5) becomes  an
algebraic equation of degree $N+1$ whose analytical solution is in general not available.  \par
Now, we expand $f$ in
powers of $\epsilon$ as
$$f=f^{(0)}+\epsilon f^{(1)}+\epsilon^2f^{(2)}+ ..., \eqno(3.10)$$
and insert this expression into (3.5). Comparing the coefficients of $\epsilon^n (n=0, 1, ...)$, we
 obtain a system of algebraic equations for $f^{(n)}$, the first two of which read
$${f^{(0)}}^2+(z_1-1)f^{(0)}=0, \eqno(3.11a)$$
$$2f^{(0)}f^{(1)}+(z_1-1)f^{(1)}=0, \eqno(3.11b)$$
$$(z_1-1)f^{(2)}-\lambda_1^2z_1=0. \eqno(3.11c)$$
The above equations can be solved immediately to obtain positive 
solutions.. Indeed,
if  $z_1\leq 1$ ( $x-c_1t-x_{10}\leq 0$), then 
$$f\sim f^{(0)}=1-z_1. \eqno(3.12a)$$ 
Substitution of this result into (3.9) yields
$$u \sim c_1z_1=c_1e^{x-c_1t-x_{10}}. \eqno(3.12b)$$
If, on the other hand, $z_1>1$(or $x-c_1t-x_{10}> 0$), then 
$$f^{(0)}=f^{(1)}=0,\qquad f\sim f^{(2)}\epsilon^2=\lambda_1^2z_1/(z_1-1)\epsilon^2, \eqno(3.13a)$$ 
$$u \sim c_1z_1^{-1}=c_1e^{-(x-c_1t-x_{10})}. \eqno(3.13b)$$
It follows from (3.12) and (3.13) that in the limit $\kappa\rightarrow 0$ (or equivalently
 $\epsilon\rightarrow 0$ by (3.3c)) $u$ has a limiting waveform 
$$ u=c_1e^{-|x-c_1t-x_{10}|}. \eqno(3.14)$$
This expression is just the $1$-peakon solution (2.6) of the CH equation with $\kappa=0$.\par
\bigskip
\leftline{\bf 4. The Peakon Limit of the $N$-Soliton Solution}\par
\leftline{\it 4.1 Formulas for determinants} \par
 The peakon limit of the general $N$-soliton solution can be taken along the lines of the
$1$-soliton case. However, the calculation involved is quite formidable. To perform the
calculation in an effective manner,
we first define the following determinants which are closely related to the $N$-peakon solution:
\begin{align}
\Delta_n(i_1,i_2, ..., i_n)&=\left|\begin{matrix}
1&1&\cdots&1 \cr
                                            \lambda_{i_1}&\lambda_{i_2}&\cdots&\lambda_{i_n}\cr
                                            \vdots&\vdots&\ddots&\vdots \cr
                                            \lambda_{i_1}^{n-1}&\lambda_{i_2}^{n-1}&\cdots&\lambda_{i_n}^{n-1}\end{matrix}
                                            \right|^2 \notag \\
                                            &=\prod_{1\leq l<m\leq n}(\lambda_{i_l}-\lambda_{i_m})^2,\qquad (n\geq 2),
                                            \tag{4.1}
                                            \end{align}
                     $$ D_n^{(m)}=\left|\begin{matrix}A_m&A_{m+1}&\cdots&A_{m+n-1}\cr
A_{m+1}&A_{m+2}&\cdots&A_{m+n}\cr
\vdots&\vdots&\ddots&\vdots \cr
A_{m+n-1}&A_{m+n}&\cdots&A_{2(n-1)+m}\cr\end{matrix}\right|. \eqno(4.2a)$$
Here
$$A_m=\sum_{i=1}^N\lambda_i^mE_i, \eqno(4.2b)$$
$$E_i=e^{{2\over\lambda_i}t+x_{i0}},\qquad  x_{i0}={y_{i0}\over \kappa},\qquad \lambda_i={2\over c_i},
\qquad (i=1, 2, ..., N), \eqno(4.2c)$$
where in (4.2) $m$ is an arbitrary integer and $n$ is a nonnegative integer less than or equal to $N$.
For $n$ greater than $N$, $D_n^{(m)}=0$.
 We use the convention $\Delta_1(i_1)=1, D_0^{(m)}=1, D_1^{(m)}=A_m$.
The quantity $\Delta_n$ is the square of the Vandermonde determinant whereas $D_n^{(m)}$ is the determinant
of a symmetric matrix. It is a Hankel determinant. We use some properties of  Hankelians in the
following analysis. \par
Let $D_n^{(m)}(i_1,i_2,...,i_p;j_1,j_2,...,j_q)\ (i_1<i_2<...<i_p, \ j_1<j_2<...<j_q,\ 1\leq p, q<N)$ be a determinant 
which is obtained from $D_n^{(m)}$ by deleting rows $i_i, i_2, ..., i_p$ and columns $j_1, j_2, ..., j_q$, respectively.
Then, the following Jacobi formula holds which will play a central role in the present analysis:$^{26)}$
$$D_{n+2}^{(m)}D_{n+2}^{(m)}(1,n+2; 1,n+2)=D_{n+2}^{(m)}(1; 1)D_{n+2}^{(m)}(n+2; n+2)-D_{n+2}^{(m)}(1; n+2)D_{n+2}^{(m)}(n+2; 1).
\eqno(4.3)$$
By virtue of the definition (4.2), we see that
$$D_{n+2}^{(m)}(1,n+2; 1,n+2)=D_{n}^{(m+2)}, \eqno(4.4a)$$
$$D_{n+2}^{(m)}(1; 1)=D_{n+1}^{(m+2)},\eqno(4.4b)$$
$$D_{n+2}^{(m)}(n+2; n+2)=D_{n+1}^{(m)},\eqno(4.4c)$$
$$D_{n+2}^{(m)}(1; n+2)=D_{n+2}^{(m)}(n+2; 1)=D_{n+1}^{(m+1)}. \eqno(4.4d)$$
Hence,  (4.3) can be rewritten in the form
$$D_{n+2}^{(m)}D_{n}^{(m+2)}=D_{n+1}^{(m+2)}D_{n+1}^{(m)}-\left(D_{n+1}^{(m+1)}\right)^2. \eqno(4.5)$$
The determinant $D_n^{(m)}$  has an alternative expression in the form of a finite sum$^{21)}$ 
$$D_n^{(m)}=\sum_{1\leq i_1<i_2<...<i_n\leq N}\Delta_n(i_1,i_2, ..., i_n)(\lambda_{i_1}\lambda_{i_2}...\lambda_{i_n})^m
E_{i_1}E_{i_2}...E_{i_n},\ (n=1, 2, ..., N). \eqno(4.6)$$
This formula  is very useful
in rewriting the tau-functions. 
Note that $D_n^{(m)}\ (n=1, 2, ..., N)$ are positive definite since $\Delta_n>0$ and $ E_i>0\ (i=1, 2, ..., N)$.
We give a simple proof of (4.6) in Appendix A. \par
\leftline{\it 4.2 The peakon limit of the N-soliton solution}\par
 We first shift the phase variables as $\xi_i\rightarrow \xi_i-\phi_i (i=1, 2, ..., N)$ in (2.2) and take the
peakon limit $\kappa k_i\rightarrow 1$ with fixed $c_i (i=1, 2, ..., N)$.
Using (3.2) and the asymptotic $e^{\gamma_{ij}}\sim(\lambda_i-\lambda_j)^2\kappa^4/16$,
the leading-order asymptotics of the tau-functions $f_1$ and $f_2$ can be written in the form
$$f_1\sim1+\sum_{n=1}^N\epsilon^{n(n-1)}\left({f_1\over f_2}\right)^n
\sum_{1\leq i_1<i_2<...<i_n\leq N}\Delta_n(i_1,i_2, ..., i_n)z_{i_1}z_{i_2}...z_{i_n},\eqno(4.7a)$$
$$f_2\sim1+\sum_{n=1}^N\epsilon^{n(n+1)}\left({f_1\over f_2}\right)^n
\sum_{1\leq i_1<i_2<...<i_n\leq N}(\lambda_{i_1}\lambda_{i_2}...\lambda_{i_n})^2
\Delta_n(i_1,i_2, ..., i_n)z_{i_1}z_{i_2}...z_{i_n},\eqno(4.7b)$$
where 
$$z_i=e^{x-c_it-x_{i0}}=e^xE_i^{-1}, \qquad (i=1, 2, ..., N). \eqno(4.7c)$$
 Furthermore, to compare 
the limiting form of $u$ resulting from the peakon limit with 
the $N$-peakon solution given by ref. 21), we shift the phase constant appropriately, so that
$$z_i \rightarrow {\prod_{i=1}^N\lambda_i^2\over 2\prod_{j=1\atop(j\not=i)}^N(\lambda_i-\lambda_j)^2}{z_i\over\lambda_i^2},
\qquad (i=1, 2, ..., N). \eqno(4.8)$$
We substitute (4.8) into (4.7) and use (4.6) to modify them into the form
$$f_1\sim 1+\sum_{n=1}^N\epsilon^{n(n-1)}\left({f_1\over f_2}\right)^nd_ne^{nx}D_{N-n}^{(2)},\eqno(4.9a)$$
$$f_2\sim 1+2\sum_{n=1}^N\epsilon^{n(n+1)}\left({f_1\over f_2}\right)^nd_{n+1}e^{nx}D_{N-n}^{(0)},\eqno(4.9b)$$
where the positive coefficients $d_n$ are defined by
$$d_n=d_n(t)={\prod_{i=1}^N\lambda_i^{2(n-1)}\over 2^n\Delta_N\prod_{i=1}^NE_i},
\qquad (n=1, 2, ..., N). \eqno(4.9c)$$
Thus,  equation corresponding to (3.5) becomes an algebraic equation of degree $N+1$
$$\sum_{n=0}^{N+1}\epsilon^{n(n-1)}h_nf^{N-n+1}+O(\epsilon^{N(N+1)+1})=0. \eqno(4.10a)$$
Here, as in the $1$-soliton case $f=f_2/f_1$ and the coefficients $h_n$ are defined by the relations
$$h_0=1,\eqno(4.10b)$$
$$h_n=d_ne^{(n-1)x}\left(e^xD_{N-n}^{(2)}-2D_{N-n+1}^{(0)}\right), \qquad (n=1, 2, ..., N),\eqno(4.10c)$$
$$h_{N+1}=-{\prod_{i=1}^N\lambda_i^{2N}e^{Nx}\over 2^N\Delta_N\prod_{i=1}^NE_i}. \eqno(4.10d)$$
To evaluate $u$ using (3.7), we need the derivatives $f_t$ and $f_x$. They are derived simply from (4.10) by 
differentiation. Explicitly
$$f_t \sim -{\sum_{n=1}^{N+1}\epsilon^{n(n-1)}h_{n,t}f^{N-n+1}\over
\sum_{n=0}^{N}\epsilon^{n(n-1)}(N-n+1)h_nf^{N-n}},\eqno(4.11a)$$
$$f_x \sim -{\sum_{n=1}^{N+1}\epsilon^{n(n-1)}h_{n,x}f^{N-n+1}\over
\sum_{n=0}^{N}\epsilon^{n(n-1)}(N-n+1)h_nf^{N-n}}.\eqno(4.11b)$$
It follows from (4.2c) and (4.10) that
$$h_{n,x}=(n-1)h_n+d_ne^{nx}D_{N-n}^{(2)}, \qquad (n=1, 2, ..., N),\eqno(4.12a)$$
$$h_{N+1,x}=Nh_{N+1},\eqno(4.12b)$$
$$h_{n,t}=-(\sum_{i=1}^Nc_i)h_n+d_ne^{(n-1)x}\left(e^xD_{N-n,t}^{(2)}-2D_{N-n+1,t}^{(0)}\right),\qquad (n=1, 2, ..., N),\eqno(4.13a)$$
$$h_{N+1,t}=-\left(\sum_{i=1}^Nc_i\right)h_{N+1}. \eqno(4.13b)$$
The expression of $u$ in terms of $f$  now follows from (3.7) and (4.10)-(4.13). 
After an elementary calculation, we find that
$$u \sim -{(\sum_{i=1}^Nc_i)f^{N}+\sum_{n=1}^N\epsilon^{n(n-1)}d_ne^{(n-1)x}\left(e^xD_{N-n,t}^{(2)}-2D_{N-n+1,t}^{(0)}\right)f^{N-n}
\over f^N+\sum_{n=1}^N\epsilon^{n(n-1)}d_ne^{nx}D_{N-n}^{(2)}f^{N-n}}.\eqno(4.14)$$
The unknown $f$ is  obtained by solving the algebraic equation (4.10). 
The procedure for constructing
 solution can now be carried out straightforwardly. Here, we 
summarize  the result. \par
We seek the series solution of the form
$$f=f^{(0)}+\epsilon^2 f^{(2)}+\epsilon^4 f^{(4)} + ..., \eqno(4.15)$$
under the conditions $f^{(n)}>0 \ (n=0, 2, 4, ...)$. 
Note that the odd powers of $\epsilon$ have been dropped in the above expansion. An inspection shows that
this can be justified since eq. (4.10) includes only the even powers of $\epsilon$. See also the $1$-peakon case in \S 3.
We  find that  the leading-order asymptotic of the positive solution of equation (4.10) is expressed simply as
$$f\sim f^{(2n-2)}\epsilon^{2(n-1)}=-{h_n\over h_{n-1}}\epsilon^{2(n-1)}, \eqno(4.16a)$$
if the following inequalities hold for $h_n \ (n=1, 2, ..., N+1)$ 
$$h_1>0,\ h_2>0, ...,\ h_{n-1}>0,\ h_n\leq 0,\ h_{n+1}<0, ...,\ h_{N+1}<0. \eqno(4.16b)$$ 
To obtain (4.16a), we assume that $f$ has a leading-order asymptotic of the form $f\sim\epsilon^{2m}f^{(2m)}$ and
subsitute this into (4.10).  Then, (4.10) is expanded in powers of $\epsilon$ as
$$\left[h_m\{f^{(2m)}\}^{N-m+1}+h_{m+1}\{f^{(2m)}\}^{N-m}\right]\epsilon^{m(2N-m+1)}+O(\epsilon^{m(2N-m+1)+2})=0. \eqno(4.17)$$
By taking the coefficient of $\epsilon^{m(2N-m+1)}$ zero, we find that $f^{(2m)}=-h_{m+1}/h_m \ (h_m\not=0)$. According to (4.16b) and
the requirement $f^{(2m)}>0$, the integer $m$ is determined uniquely as $m=n-1$, which immediately leads to (4.16a).
Note from (4.10c) that the coordinate $x=x_n$ giving rise to the equality $h_n=0$ can be
specified as
$$x_n={\rm ln}\left[{2D_{N-n+1}^{(0)}\over D_{N-n}^{(2)}}\right],\qquad  (n=1, 2, ..., N), \eqno(4.18)$$
and $h_n\geq 0$ for $x\geq x_n \ (n=1, 2, ..., N)$.
An important consequence deduced from (4.18) is the notable inequalities 
$$x_{j-1}<x_{j},\qquad (j=1, 2, ..., N+1), \eqno(4.19)$$
where we have used the convention $x_0=-\infty$ and $x_{N+1}=+\infty$.
In fact, we substitute  (4.18) into (4.19) to rewrite them in the following alternative forms
$$D_{N-j+1}^{(0)}D_{N-j+1}^{(2)}-D_{N-j+2}^{(0)}D_{N-j}^{(2)}>0, \qquad (j=1, 2, ..., N+1). \eqno(4.20)$$
The left-hand side of (4.20) equals to $\left(D_{N-j+1}^{(1)}\right)^2$ by Jacobi's
identity (4.5) with $m=0$ and $n=N-j$ and consequently it always has a positive value, which proves (4.20).
We see from (4.10c), (4.18) and (4.19) that under the inequalities (4.16b),
  $x$ must lie in the interval $x_{n-1}<x\leq x_n (n=1, 2, ..., N+1)$.  \par
Let $u_n$ be the waveform of $u(x,t)$ for $x$  in the interval $x_{n-1}<x\leq x_{n} (n=1, 2, ..., N)$ at 
any instant $t$. 
 We substitute (4.16) into (4.14) and  
 see that both the denominator and numerator of (4.14) have a leading-order asymptotic of order $\epsilon^{(n-1)(2N-n)}$.
 Consequently, 
  the limit  exists when $\epsilon$ tends to zero. It turns out that  $u_n$
has a limiting waveform given by
$$u_n={G_n\over F_n}, \eqno(4.21a)$$
where
$$F_n=-d_{n-1}D_{N-n+1}^{(2)}h_n+d_ne^xD_{N-n}^{(2)}h_{n-1}, \eqno(4.21b)$$
$$G_n=d_{n-1}e^{-x}\left(e^xD_{N-n+1,t}^{(2)}-2D_{N-n+2,t}^{(0)}\right)h_n
-d_n\left(e^xD_{N-n,t}^{(2)}-2D_{N-n+1,t}^{(0)}\right)h_{n-1}.\eqno(4.21c)$$
Inserting $h_n$ from (4.10c),  $F_n$  becomes
$$F_n=2d_{n-1}d_ne^{(n-1)x}\left(D_{N-n+1}^{(2)}D_{N-n+1}^{(0)}-D_{N-n}^{(2)}D_{N-n+2}^{(0)}\right). \eqno(4.22)$$
We use the formula (4.5) with $m=0$ and $n$ replaced by $N-n$ in (4.22). Then, $F_n$ simplifies to
$$F_n=2d_{n-1}d_ne^{(n-1)x}\left[D_{N-n+1}^{(1)}\right]^2. \eqno(4.23)$$
By a similar calculation,  $G_n$  is transformed to
$$G_n=2d_{n-1}d_ne^{(n-1)x}\Bigg[{e^x\over 2}\left\{D_{N-n+1,t}^{(2)}D_{N-n}^{(2)}-D_{N-n,t}^{(2)}D_{N-n+1}^{(2)}\right\}$$
$$+D_{N-n,t}^{(2)}D_{N-n+2}^{(0)}+D_{N-n+1,t}^{(0)}D_{N-n+1}^{(2)}
-D_{N-n+1,t}^{(2)}D_{N-n+1}^{(0)}-D_{N-n+2,t}^{(0)}D_{N-n}^{(2)}$$
$$+2e^{-x}\left\{D_{N-n+2,t}^{(0)}D_{N-n+1}^{(0)}-D_{N-n+1,t}^{(0)}D_{N-n+2}^{(0)}\right\}\Bigg]. \eqno(4.24)$$
The following identities among determinants are verified with the aid of  the Jacobi identity (see Appendix B):
$$D_{N-n+1,t}^{(2)}D_{N-n}^{(2)}-D_{N-n,t}^{(2)}D_{N-n+1}^{(2)}=2D_{N-n+1}^{(1)}D_{N-n}^{(3)},\eqno(4.25a)$$
$$D_{N-n+2,t}^{(0)}D_{N-n+1}^{(0)}-D_{N-n+1,t}^{(0)}D_{N-n+2}^{(0)}=2D_{N-n+2}^{(-1)}D_{N-n+1}^{(1)},\eqno(4.25b)$$
$$D_{N-n,t}^{(2)}D_{N-n+2}^{(0)}-D_{N-n+1,t}^{(2)}D_{N-n+1}^{(0)}=-2D_{N-n+2}^{(-1)}(1; 2)D_{N-n+1}^{(1)},\eqno(4.25c)$$
$$D_{N-n+1,t}^{(0)}D_{N-n+1}^{(2)}-D_{N-n+2,t}^{(0)}D_{N-n}^{(2)}=2D_{N-n+2}^{(-1)}(1; 2)D_{N-n+1}^{(1)}. \eqno(4.25d)$$
Substituting  (4.25) into (4.24), we can  reduce (4.24) considerably, giving rise to 
$$G_n=2d_{n-1}d_ne^{(n-1)x}D_{N-n+1}^{(1)}\left[e^xD_{N-n}^{(3)}+4e^{-x}D_{N-n+2}^{(-1)}\right].\eqno(4.26)$$
The expression of $u_n$ now follows from (4.21a), (4.23) and (4.26). It is expressed compactly in terms of
the Hankel determinants as
$$u_n={e^xD_{N-n}^{(3)}+4e^{-x}D_{N-n+2}^{(-1)}\over D_{N-n+1}^{(1)}},\qquad (n=1, 2, ..., N).\eqno(4.27)$$
Note that when $n=1$, (4.27) reduces to $u_1=e^xD_{N-1}^{(3)}/D_N^{(1)}$ due to the relation $D_{N+1}^{(-1)}=0$
which represents the waveform of $u$ in the range $x\leq x_1$. Using (4.6), this expression can be written in the form
$$u_1=\sum_{i=1}^Nb_iz_i, \eqno(4.28a)$$
with
$$b_i=\prod_{j=1}^N\lambda_j^2\prod_{j=1\atop(j\not=i)}^N(\lambda_i-\lambda_j)^{-2}\lambda_i^{-3},\qquad (i=1, 2, ..., N),
\eqno(4.28b)$$
where $\lambda_i$ and $z_i$ are defined respectively by (4.2c) and (4.7c).
\par
For $x$ in the range $x_N\leq x$, we find  that $u=u_{N+1}$ has a particularly simple form
$$u_{N+1}=2e^{-x}D_{1,t}^{(0)}=2\sum_{i=1}^Nc_iz_i^{-1}, \eqno(4.29)$$
where we have used $D_1^{(0)}=A_0$ , (4.2b) and (4.7c). \par
\leftline{\it 4.3 Comparison with the N-peakon solution}\par
 Here, we compare the expression of the peakon limit arising from the $N$-soliton solution with the
$N$-peakon solution given by Beals et al.$^{21)}$ 
If we identify the quantities $\Delta_n^m\ (m>0)$ and $\tilde\Delta_n^0$ introduced in ref. 21
with $D_n^{(m)}\ (m>0)$ and $D_n^{(0)}$, respectively, then we can write the latter solution in the form
$$u(x,t)=\sum_{i=1}^Nm_i(t)e^{-|x-x_i(t)|},\eqno(4.30a)$$
$$m_i={2D_{N-i+1}^{(0)}D_{N-i}^{(2)}\over D_{N-i+1}^{(1)}D_{N-i}^{(1)}},\qquad (i=1, 2, ..., N),\eqno(4.30b)$$
$$x_i={\rm ln}\left[{2D_{N-i+1}^{(0)}\over D_{N-i}^{(2)}}\right],\qquad (i=1, 2, ..., N). \eqno(4.30c)$$
When $x$ lies in the interval $x_{n-1}<x\leq x_n$, then $u=u_n$ takes the form
\begin{align}
u_n
&=\sum_{i=1}^{n-1}m_ie^{-(x-x_i)}+\sum_{i=n}^Nm_ie^{-(x_i-x)}\notag  \\
&=e^{-x}\sum_{i=N-n+1}^{N-1}{4\left(D_{i+1}^{(0)}\right)^2\over D_{i+1}^{(1)}D_i^{(1)}}+e^{x}\sum_{i=0}^{N-n}{\left(D_{i}^{(2)}\right)^2
\over D_{i+1}^{(1)}D_i^{(1)}}. \tag{4.31}
\end{align}
  The following formulas are particularly useful to simplify (4.31):
  $$\sum_{i=0}^{N-n}{\left(D_{i}^{(2)}\right)^2\over D_{i+1}^{(1)}D_i^{(1)}}={D_{N-n}^{(3)}\over D_{N-n+1}^{(1)}},
  \qquad (n=0, 1, ..., N),\eqno(4.32a)$$
$$\sum_{i=N-n+1}^{N-1}{\left(D_{i+1}^{(0)}\right)^2\over D_{i+1}^{(1)}D_i^{(1)}}={D_{N-n+2}^{(-1)}\over D_{N-n+1}^{(1)}},
  \qquad (n= 2, 3, ..., N+1).\eqno(4.32b)$$
To prove (4.32a), we write Jacobi's formula (4.5) with $m=1$ and $n=i-1$. It reads
$$\left(D_i^{(2)}\right)^2=D_i^{(3)}D_i^{(1)}-D_{i+1}^{(1)}D_{i-1}^{(3)}. \eqno(4.33)$$
Dividing (4.33) by $D_{i+1}^{(1)}D_i^{(1)}$ and summing up the resultant expression from $i=1$ to $i=N-n$, we obtain
(4.32a) by taking account of the relations 
$D_0^{(m)}=1\ (m=1, 2, 3)$. By a similar calculation using (4.5) with $m=-1$ and $n=i$,
formula (4.32b) follows immediately upon noting the relation $D_{N+1}^{(-1)}=0$. If we introduce (4.32) 
into (4.31), we find that the resultant expression coincides with the peakon limit of the $N$-soliton solution  (4.27). \par
Last, for $x_N\leq x$, (4.30) becomes
$$u=u_{N+1}=4e^{-x}\sum_{i=0}^{N-1}{\left(D_{i+1}^{(0)}\right)^2\over D_{i+1}^{(1)}D_i^{(1)}}.\eqno(4.34)$$
Using (4.32b) with $n=N+1$, we can recast (4.34) to
$$u_{N+1}=4e^{-x}D_1^{(-1)}=4e^{-x}\sum_{i=1}^N\lambda_i^{-1}E_i,\eqno(4.35)$$
which is in agreement with (4.29) by (4.2c) and (4.7c).
In conclusion, we have completed the proof that the $N$-soliton solution
of the CH equation reduces to the $N$-peakon solution of the dispersionless CH equation
in the peakon limit. \par
\bigskip
\leftline{\bf 5. Concluding Remarks}\par
 We have developed a novel limiting procedure for recovering the nonanalytic
$N$-peakon solution from the analytic $N$-soliton solution of the CH equation. In the process, the
asymptotic analysis has been performed which expands all the wave parameters in
powers of the small dispersion parameter. 
We have employed the key identity, i.e. Jacobi's formula (4.5)
for determinants by which the significant part of the calculations has been carried out in quite a
transparent manner. In view of this formula, we were also able to obtain the new
formulas (4.27) and (4.28) which represent the waveform of the $N$-peakon solution
in a specified interval of the space variable. 
Our approach is quite different from
 that used by Beals et al$^{20,21)}$ which relies on the classical
moment problem originally studied by Stieltjes.
In a future work, we have a plan to apply our
method to the $N$-soliton solution$^{27,28)}$ of the Degasperis-Procesi (DP) equation to
show that it recovers the $N$-peakon solution already given by Lundmark and Szmigielski.$^{29,30)}$
 While both the CH and DP equations have a different mathematical structure, the
$N$-soliton solution of the DP equation can be written in a parametric form just like (2.1) and (2.2),
the only difference being the structure of the tau-functions $f_1$ and $f_2$. 
This observation will make it possible to perform the peakon limit of the $N$-soliton
solution in a manner similar to that developed here for the CH equation.
\par
\bigskip
\leftline{\bf Acknowledgement}\par
 The author wishes to thank Dr Allen Parker for kind inspection of the manuscript and
useful comments. \par
\newpage
\leftline{\bf Appendix A: Proof of (4.6)}\par
 We introduce the $n\times N$ matrix $P$ and the $N\times n$ matrix $Q$
$$P=\left(\begin{matrix}1&1&\cdots&1 \cr
                                            \lambda_{1}&\lambda_{2}&\cdots&\lambda_{N}\cr
                                            \vdots&\vdots&\ddots&\vdots \cr
                                            \lambda_{1}^{n-1}&\lambda_{2}^{n-1}&\cdots&\lambda_{N}^{n-1}\end{matrix}
                                            \right),             \eqno(A\cdot1)$$

$$Q=\left(\begin{matrix}\lambda_1^mE_1&\lambda_1^{m+1}E_1&\cdots& \lambda_1^{m+n-1}E_1\cr
                                            \lambda_2^mE_2&\lambda_2^{m+1}E_2&\cdots&\lambda_2^{m+n-1}E_2\cr
                                            \vdots&\vdots&\ddots&\vdots \cr
                                            \lambda_N^mE_N&\lambda_N^{m+1}E_N&\cdots&\lambda_N^{m+n-1}E_N\end{matrix}
                                            \right).             \eqno(A\cdot2)$$
   We calculate ${\rm det}(PQ)$ by two ways. First,  multiplying $P$ and $Q$ and using (4.2), we obtain 
   $${\rm det}(PQ)=D_n^{(m)}. \eqno (A\cdot3)$$
    On the other hand, the determinant is expanded by means of the Binet-Cauchy formula as$^{31)}$
     \begin{align}
     {\rm det}(PQ)
        &=\sum_{1\leq i_1<i_2<...<i_n\leq N} \left|\begin{matrix}1&1&\cdots&1 \cr
                                            \lambda_{i_1}&\lambda_{i_2}&\cdots&\lambda_{i_n}\cr
                                            \vdots&\vdots&\ddots&\vdots \cr
                                            \lambda_{i_1}^{n-1}&\lambda_{i_2}^{n-1}&\cdots&\lambda_{i_n}^{n-1}\end{matrix}
                                            \right| 
                                             \left|\begin{matrix}\lambda_{i_1}^mE_{i_1}&\lambda_{i_1}^{m+1}E_{i_1}&\cdots
                                             & \lambda_{i_1}^{m+n-1}E_{i_1}\cr
                                            \lambda_{i_2}^mE_{i_2}&\lambda_{i_2}^{m+1}E_{i_2}&\cdots&\lambda_{i_2}^{m+n-1}E_{i_2}\cr
                                            \vdots&\vdots&\ddots&\vdots \cr
                                            \lambda_{i_n}^mE_{i_n}&\lambda_{i_n}^{m+1}E_{i_n}&\cdots&\lambda_{i_n}^{m+n-1}E_{i_n}\end{matrix} 
                                            \right|   \notag \\
    &= \sum_{1\leq i_1<i_2<...<i_n\leq N}\Delta_n(i_1,i_2,...,i_n)
    (\lambda_{i_1}\lambda_{i_2}...\lambda_{i_n})^mE_{i_1}E_{i_2}...E_{i_n}. \tag{A$\cdot$4}
    \end{align}
    In passing to the last line of (A$\cdot$4), we have extracted the factor $\lambda_{i_l}^mE_{i_l}$
    from the $l$th row \ $(l=1, 2, ..., n)$ and 
    used (4.1). Formula (4.6) follows immediately from (A$\cdot$3) and (A$\cdot$4).  \par
    \bigskip
    \leftline{\bf Appendix B: Formulas for Determinants}\par
    Here, we prove (4.25). We define the quantity $D$ by the relation
    $$ D=D_{n+1,t}^{(2)}D_n^{(2)}-D_{n+1}^{(2)}D_{n,t}^{(2)}. \eqno(B\cdot1)$$
    We differentiate $D_n^{(2)}$ by $t$ with use of the relation $dA_m/dt=2A_{m-1}$ which is a consequence of (4.2b).
    The differentiation of the $j$th column of  $D_n^{(2)}$ is then proportional to the $(j-1)$th column. Thus, it vanishes
    identically except for $j=1$. Hence
    $$ D_{n,t}^{(2)}=2\left|\begin{matrix}A_1&A_{3}&\cdots&A_{n+1}\cr
A_{2}&A_{4}&\cdots&A_{n+2}\cr
\vdots&\vdots&\ddots&\vdots \cr
A_{n}&A_{n+2}&\cdots&A_{2n}\end{matrix}\right|. \eqno(B\cdot2)$$
Using (B$\cdot$2) and the relations (4.4), (B$\cdot$1) becomes
$$D=2D_{n+2}^{(1)}(n+2;2)D_{n+2}^{(1)}(n+1,n+2; 1,n+2)
- 2D_{n+2}^{(1)}(n+2;1)D_{n+2}^{(1)}(n+1,n+2; 2,n+2). \eqno(B\cdot3)$$
It follows by using Jacobi's formula (4.3) and a fact $D_{n+2}^{(1)}>0$ \ (see (4.6)) that
$$D_{n+2}^{(1)}(n+1,n+2; 1,n+2)$$
$$={1\over D_{n+2}^{(1)}}\left[D_{n+2}^{(1)}(n+1; 1) D_{n+2}^{(1)}(n+2; n+2)
-D_{n+2}^{(1)}(n+1; n+2) D_{n+2}^{(1)}(n+2; 1)\right], \eqno(B\cdot4)$$
 $$D_{n+2}^{(1)}(n+1,n+2; 2,n+2)$$
$$={1\over D_{n+2}^{(1)}}\left[D_{n+2}^{(1)}(n+1; 2) D_{n+2}^{(1)}(n+2; n+2)
-D_{n+2}^{(1)}(n+1; n+2) D_{n+2}^{(1)}(n+2; 2)\right]. \eqno(B\cdot5)$$ 
Substituting (B$\cdot$4) and (B$\cdot$5) into (B$\cdot$3), we obtain
$$ D={2D_{n+2}^{(1)}(n+2; n+2)\over D_{n+2}^{(1)}} 
\left[ D_{n+2}^{(1)}(n+1; 1)D_{n+2}^{(1)}(n+1; 2)- D_{n+2}^{(1)}(n+1; 2)D_{n+2}^{(1)}(n+1; 1) \right]. \eqno(B\cdot6)$$
We apply Jacobi's formula to the right-hand side of (B$\cdot$6) and use the relations (4.4). 
Then, (B$\cdot$6) is recast to a simple form
$$ D=2D_{n+1}^{(1)}D_n^{(3)}. \eqno(B\cdot7)$$
  Formula (4.24a) is a consequence of (B$\cdot$1) and (B$\cdot$7) with $n$ replaced by $N-n$. Formula (4.24b) follows from (4.24a).
  To show this, we may simply replace $n$ by $n-1$ and $E_i$ by $\lambda_i^{-2}E_i (i=1, 2, ..., N)$, respectively in (4.24a).\par
  By repeating the similar calculation, we can show that
  \begin{align}
  D_{n,t}^{(2)}D_{n+2}^{(0)}- D_{n+1,t}^{(2)}D_{n+1}^{(0)} 
  & =-2D_{n+3}^{(-1)}(1,n+3; 2,n+3)D_{n+3}^{(-1)}(n+2,n+3; 1,2) \notag \\
  &=-2D_{n+2}^{(-1)}(1; 2)D_{n+1}^{(1)}. \tag{B$\cdot$8}
  \end{align}
  If we replace $n$ by $N-n$ in (B$\cdot$8), we obtain (4.25c).  Formula (4.25d) is derived by the same way. \par
  \newpage
  \leftline{\bf References}
    \begin{enumerate}[1)]
  \item
   R. Camassa  and D.D. Holm: Phys. Rev. Lett. {\bf 71} (1993) 1661.
  \item
   R. Camassa, D. D. Holm  and J. Hyman: Adv. Appl. Mech. {\bf 31} (1994) 1.
   \item
   R. S. Johnson: J. Fluid Mech. {\bf 455} (2002) 63.
   \item
   H. R. Dullin, G. A. Gottwald and D. D. Holm: Physica D {\bf 190} (2004) 1.                                                                   
  \item
  A. Parker: Proc. R. Soc. London Ser. A {\bf 460} (2004) 2929.
  \item
  J. Schiff: Physica D {\bf 121} (1998) 24.
  \item
  M. S. Alber and Yu. Fedorov: J. Phys. A: Math. Gen. {\bf 33} (2000) 8409.
  \item
  M. S. Alber and Yu. Fedorov:  Inverse Probl. {\bf 17} (2001) 1017.
  \item
  R. S. Johnson:  Proc. R. Soc. London Ser. A {\bf 459} (2003) 1687.
  \item
  Y. Li Y and J. E. Zhang:  Proc. R. Soc. London Ser. A {\bf 460} (2004) 2617.
  \item
  Y. Matsuno: J. Phys. Soc. Jpn. {\bf 74} (2005) 1983.
  \item
  Y. Li Y: J. Nonlinear Math. Phys. {\bf 12} (Suppl. 1) (2005) 466.
  \item
  A. Parker:  Proc. R. Soc. London Ser. A {\bf 461} (2005) 3611.
  \item
  A. Parker:  Proc. R. Soc. London Ser. A {\bf 461} (2005) 3893.
  \item
  A. Parker: Inverse Probl. {\bf 22} (2006) 599.
  \item
  R. A. Kraenkel  and A. Zenchuk: J. Phys. A: Math. Gen. {\bf 32} (1999) 4733.
  \item
  M. C. Ferreira, R. A. Kraenkel and A. Zenchuk:  J. Phys. A: Math. Gen. {\bf 32} (1999) 8665.
 \item
  H. H. Dai  and Y. Li:  J. Phys. A: Math. Gen. {\bf 38} (2005) L685.
  \item
  Y. Matsuno: Phys. Lett. A{\bf 359} (2006) 451. 
   \item
  R. Beals, D. H. Sattinger  and J. Szmigielski:  Inverse Probl. {\bf 15} (1999) L1.
  \item
  R. Beals, D. H. Sattinger  and J. Szmigielski:  Adv. Math. {\bf 154} (2000) 229.
  \item
A. Parker: preprint (2006).
\item
  Y. A. Li and P. J. Olver: Discrete Contin. Dynam. Syst. {\bf A3} (1997) 419.
  \item                                            
  Y. A. Li and P. J. Olver: Discrete Contin. Dynam. Syst. {\bf A4} (1997) 159.
  \item
  A. Parker  and Y. Matsuno: J. Phys. Soc. Jpn. {\bf 75} (2006) 124001.
  \item
  R. Vein  and P. Dale: {\it Determinants and Their Applications in Mathematical Physics} (Springer, 1999).
    \item
  Y. Matsuno: Inverse Probl. {\bf 21} (2005) 1553.
  \item
  Y. Matsuno: Inverse Probl. {\bf 21} (2005) 2085.
   \item
   H. Lundmark and J. Szmigielski:  Inverse Probl. {\bf 19} (2003) 1241.
      \item
    H. Lundmark and J. Szmigielski:  Int. Math. Res. Paper {\bf 2005} (2005) 53.
   \item
  F. R. Gantmacher: {\it The Theory of Matrices} Vol. I (Chelsea, 1977).

\end{enumerate}
\end{document}